# Interfacial Coupling Reshapes Nonlinear Polarization in a Spin-Defect van der Waals Heterostructure

L. Dalifoski[a], E. S. Mayner[b], P. Singh[a*]

[a]*Laboratory for Fundamental BioPhotonics (LBP), Institute of Bioengineering (IBI), School of Engineering (STI), École Polytechnique Fédérale de Lausanne (EPFL), CH-1015 Lausanne, Switzerland;*

[b]*Laboratory of Nanoscale Biology (LBEN), Institute of Bioengineering (IBI), School of Engineering (STI), École Polytechnique Fédérale de Lausanne (EPFL), 1015 Lausanne, Switzerland;*

**To whom correspondence may be addressed. E-mail: poonam.singh@epfl.ch.*

**Keywords:** hexagonal boron nitride; negatively charged boron vacancy; van der Waals heterostructure; polarization-resolved second-harmonic generation; nonlinear interfacial coupling

## Abstract

Van der Waals heterostructures offer a direct route to couple an optically active layer to a two-dimensional spin sensor, yet their coherent nonlinear response at interfaces remains poorly resolved. Here, we use polarization-resolved second harmonic-generation imaging to resolve four distinct regions of a hexagonal boron nitride (hBN)-based heterostructure: native hBN, neutron-irradiated hBN containing negatively charged boron vacancies ($V_{B^-}$), $PbI_2$ and their complete overlap. The detected intensity scales quadratically with incident power, consistent with a dominant second-order optical process. Polarization scans show that adding $PbI_2$ transforms the distorted $V_{B^-}$ SHG anisotropy into a predominantly fourfold response with a sixfold stronger SHG than native hBN alone. This non-additive nonlinear response occurs in the same layer architecture also shown to enhance defect photoluminescence, linking spin-defect sensitization to an interface-modified effective second-order susceptibility.

## Introduction

Van der Waals heterostructures combine materials with distinct optical, electronic and spin properties within a single planar architecture without conventional lattice-matching constraints [1, 2]. The constituent layers largely retain their individual electronic structures while interacting across atomically close interfaces. Depending on the band alignment and spectral overlap, photoexcitation can drive ultrafast electron or hole transfer [3], form spatially indirect interlayer excitons [4-6], or transfer excitation energy through resonant dipole-dipole coupling without net charge displacement [7]. Changes in dielectric screening, interfacial electric fields, strain, stacking and short-range hybridization can further alter exciton energies and radiative or non-radiative decay pathways [8-10]. The optical response of a heterostructure therefore emerges from several coupled processes rather than a unique microscopic pathway.

Experimental studies have tried to distinguish these processes by combining complementary measurements. Spatially resolved photoluminescence identifies donor quenching, acceptor enhancement and new interlayer-emission bands, while photoluminescence-excitation spectroscopy relates these changes to absorption in the constituent layers [5, 7]. Time-resolved photoluminescence and femtosecond transient-absorption spectroscopy determine the rates of charge and energy transfer [3, 5, 7]. Reflectance and absorption spectroscopy resolve excitonic resonances and their renormalization, whereas Raman spectroscopy identifies changes in strain, doping and interlayer vibrational coupling [4, 6, 8]. Surface-potential, photoemission and scanning-probe measurements further constrain band alignment and interfacial charge redistribution [9, 10]. Together, these methods establish the energetic and population dynamics of a heterostructure. However, they do not determine how interfacial coupling modifies its coherent nonlinear polarization. Moreover, donor quenching or acceptor enhancement can result from altered absorption, population transfer, radiative yield or non-radiative decay. Photoluminescence alone therefore rarely identifies a unique microscopic pathway.

Hexagonal boron nitride is particularly attractive for coupling interfacial photophysics to quantum sensing. Its chemical stability, atomically flat surface and wide bandgap have established hBN as an encapsulation and dielectric layer in van der Waals devices [11-13], while its surface-accessible optical emitters enable interfacial sensing [14]. While hBN can host a variety of optically active defects, neutron or ion irradiation can be used to reliably create a well-characterized class of defects known as the negatively charged boron vacancies, $V_{B^-}$. The $V_{B^-}$ centre has a spin-triplet ground state defect that displays optically detected magnetic resonance room-temperature [15, 16]. The atomically thin host places these defects within nanometres of the surface, and therefore of a target system, enabling measurements of magnetic fields, temperature, pressure, strain and paramagnetic species [16-21]. The highest point group symmetry of the $V_{B^-}$ centre is $D_{3h}$. However, strain, electric fields and Jahn-Teller distortions can lower its local symmetry to $C_{2v}$ [22-26]. Its electronic manifold originates from the in-plane dangling bonds and out-of-plane $p_z$ orbitals of the three neighbouring nitrogen atoms [22, 23, 27]. Local structural distortions can therefore modify the defect-state energies, optical selection rules and polarization-dependent response. Although the $V_{B^-}$ has already demonstrated utility in sensing, its intrinsically weak photoluminescence, caused by dipole-forbidden transitions between ground and excited state, limits practical sensitivity of the spin sensor and therefore the applicability.

This limitation was recently addressed by integrating irradiated hBN with lead iodide, $PbI_2$, as an optically active sensitizing layer [28]. The hBN-$PbI_2$ heterostructure enhanced $V_{B^-}$ photoluminescence by 5-10 (room temperature) to-45-fold (low temperature) while retaining the optically detected magnetic-resonance response. Quenching of $PbI_2$ emission and enhancement of defect emission within the overlap region, together with type-I band alignment and strong spectral overlap, support short-range interlayer energy-transfer pathways [28]. Electronic-structure calculations further indicate limited ground-state hybridization between $PbI_2$ and the defect states, while showing that the optical response of the coupled interface can be strongly modified [28]. These measurements establish the optical and spin benefits of the heterostructure, but they do not determine how coupling manifests in its coherent second-order polarization. In particular, the nonlinear responses of native hBN, irradiated $V_{B^-}$-hBN, the $PbI_2$-containing region and their complete overlap yet remain elusive. It therefore remains unclear how $PbI_2$ reorganizes the amplitude, anisotropy and effective symmetry of the nonlinear response.

Polarization-resolved second-harmonic generation (SHG) provides a complementary symmetry-sensitive readout (Figure. 1a) of buried interfaces. Here, the generated field is determined by the symmetry and tensorial second-order susceptibility of the coupled system. In layered materials such as hBN or transition-metal dichalcogenides (TMD), SHG has resolved layer parity, crystallographic orientation, stacking order and twist angle [29-33], while changes in SHG amplitude and anisotropy have revealed the effects of defects, interlayer coupling and resonant electronic structure [32-38].

Extending SHG beyond structural characterization, Chen *et al.* showed that near-field dipolar coupling in an α-perylene/$WS_2$ heterostructure produces a large SHG enhancement and converts the six-lobed response of monolayer $WS_2$ into a dipole-aligned two-lobed pattern [36]. Their results demonstrate that an optically active donor can reshape the coherent nonlinear polarization of an adjacent layer rather than only redistribute its excited-state population. Spatial SHG imaging and resonant measurements have likewise resolved local-field and interface-induced modifications in two-dimensional systems [32, 34, 35, 38, 39]. However, this approach has not been applied to a spin-defect heterostructure in which energy transfer is used to enhance quantum-sensor emission, nor has the nonlinear response been spatially decomposed into native, defect-containing, sensitizer-containing and fully coupled regions of the same stack. Polarization-resolved SHG can therefore identify whether heterostructure formation changes the effective nonlinear anisotropy or optical-dipole orientation. Here, we first benchmark polarization-resolved SHG imaging in native hBN and then apply it to a native-hBN/$V_{B^-}$-hBN/$PbI_2$/native-hBN heterostructure to resolve the nonlinear responses of its constituent and fully overlapping regions.

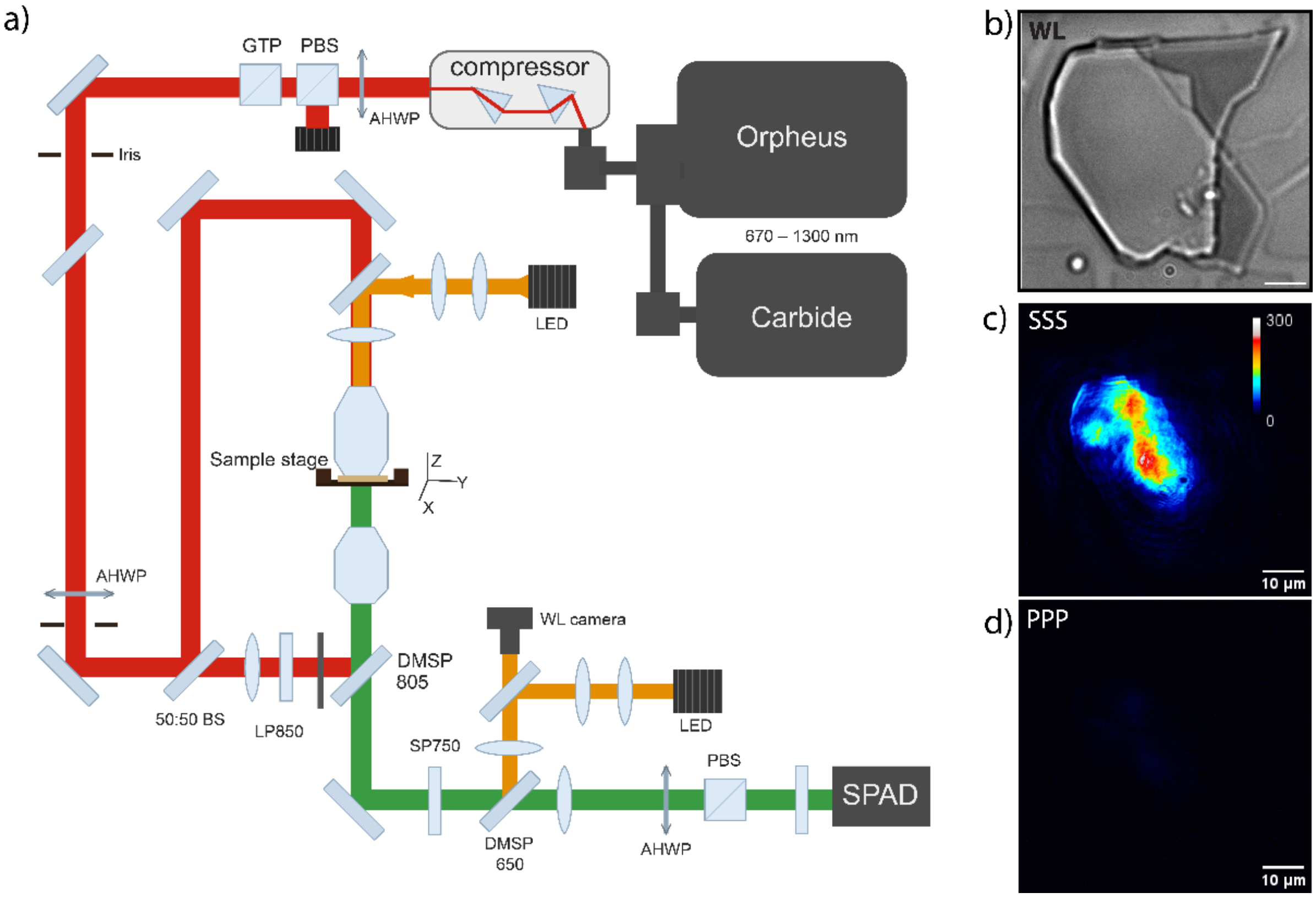

**Figure 1. Polarization-resolved SHG establishes the native-hBN reference.** (a) Schematic of the second harmonic-generation microscope, showing incident-polarization control, sample illumination, white-light imaging, and polarization-selective detection of the SHG intensity. All SHG images were acquired in reflection geometry. Here, GTP: Glan-Taylor polarizer; PBS: polarizing beam splitter; AHWP: achromatic half-wave plate; BS: beam splitter; LP: long-pass filter; SP: short-pass filter; DMSP: short-pass dichroic mirror; WL: white light; LED: light-emitting diode; and SPAD: single-photon avalanche-diode camera. (b) White-light image of two overlapping hBN flakes. (c,d) SHG intensity images of the same region acquired in the SSS and PPP polarization configurations, respectively. The first letter denotes the detected SHG polarization at 2ω, and the last two letters denote the incident fundamental polarization ω; S and P indicate polarization perpendicular and parallel to the plane of incidence, respectively. Each SHG image represents the average of 20 frames acquired per measurement and processed in Fiji. Scale bars in (b-d): 10 µm. (c,d) share a 0–300-count intensity scale.

Figure 1 establishes the polarization-resolved SHG measurement and the native-hBN reference required for this comparison. The reflection-mode microscope controls the incident linear polarization and independently selects the detected SHG polarization. The back-focal-plane illumination follows established medium-repetition-rate wide-field SH microscopy architectures, enabling parallel acquisition over an approximately 90 μm field of view [40, 41]. A white-light image identifies two overlapping native-hBN flakes (Figure 1b), while the corresponding SSS and PPP images reveal markedly different spatial intensity distributions (Figure 1c,d). Under identical acquisition conditions, the PPP counts are approximately ten times lower than the SSS counts, and the overlap region is more prominent in SSS. In the wide-field geometry used here, SSS predominantly detects in-plane nonlinear polarization. Thus, the stronger native-hBN response in SSS is consistent with the leading in-plane electric-dipole SHG response of non-centrosymmetric hBN layers, for which $D_3h$ symmetry suppresses out-of-plane susceptibility components [30, 33]. This polarization contrast demonstrates that a single polarization SHG image cannot be interpreted as a direct map of layer thickness or intrinsic susceptibility. Stacking itself can change the coherently generated SH field through symmetry, orientation and interference effects [30, 31, 33]. The native-hBN measurement therefore provides a reference against which changes associated with irradiation and heterostructure formation can be evaluated using absolute intensity, polarization anisotropy and the full angular response together.

On this basis, we compare the absolute SHG intensity, mixed-polarization contrast and full angular response of the four regions within the same heterostructure. This region-by-region strategy distinguishes scalar SHG enhancement from changes in the relative nonlinear-dipole contributions associated with irradiation-induced symmetry lowering and $PbI_2$-$V_{B-}$ interfacial coupling.

## Results and Discussion

**Polarization resolves layer-specific SHG contrast in the heterostructure:**

Polarization-resolved SHG imaging was extended from native hBN to a four-layer van der Waals heterostructure comprising bottom native hBN, neutron-irradiated $V_{B-}$-hBN, $PbI_2$ and top native hBN (Figure 2a,b). Their partial lateral overlap defines four distinct regions: native hBN, $V_{B-}$-rich hBN, native-hBN/$PbI_2$ and the complete four-layer stack. The fabrication procedure and layer definitions are described in Supplementary Section S1.1. In all four regions, $\sqrt{I_{SHG}}$ increased approximately linearly with incident power under PPP polarization (Figure 2c). For a fixed laser repetition rate of 400kHz used here, this corresponds to a quadratic dependence of $I_{SHG}$ on pulse energy and is consistent with the detected 475 nm signal being dominated by a second-order process over the measured range. The complete overlap exhibits the steepest dependence and the largest absolute PPP response. At each measured average pump power at sample, the mean SHG intensity of complete overlap exceeded compared to the constituent-regions.

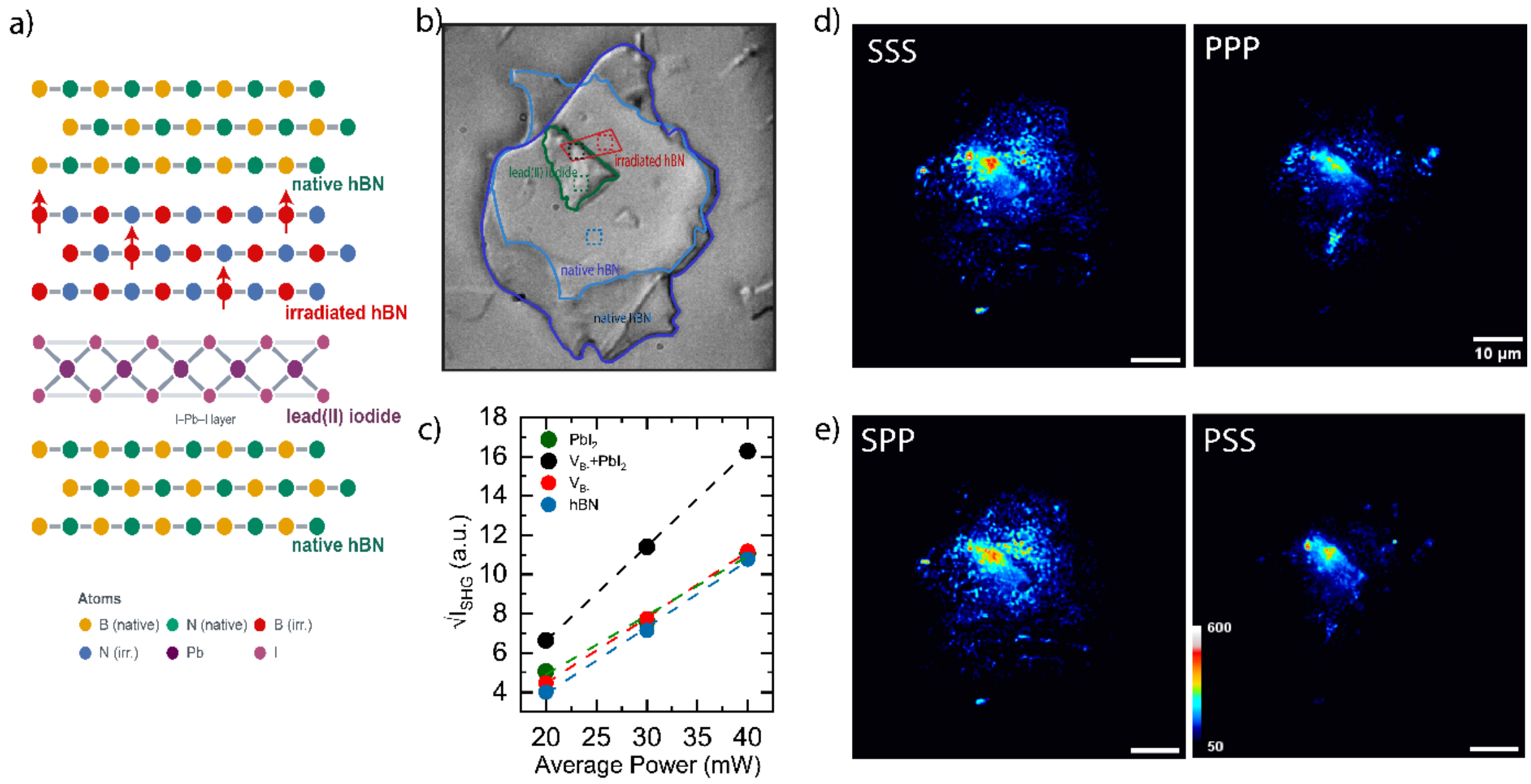

**Figure 2. Polarization resolves layer-specific SHG contrast in the hBN/$PbI_2$ heterostructure.** (a) Schematic of the four-layer van der Waals heterostructure. (b) White-light image of the four-flake heterostructure comprising bottom native hBN (dark blue), $PbI_2$ (green), neutron-irradiated $V_{B^-}$-hBN (red) and top native hBN (light blue). Colored outlines indicate the flake boundaries. Colored squares mark the equal-area regions of interest used in (c) and follow the same color scheme as the plotted data. (c) Square root of the SHG intensity, $\sqrt{I_{SHG}}$, as a function of average excitation power in the PPP configuration for native hBN, $V_{B^-}$-rich hBN, the native-hBN/$PbI_2$ region and the complete overlap. Lines are linear fits. The linear dependence of $\sqrt{I_{SHG}}$ is consistent with a quadratic scaling of $I_{SHG}$ with average pump power at sample. For 40mW, this corresponds to a pulse energy of 0.1µJ. (d) SHG intensity maps of the same field of view acquired in the SSS and PPP configurations, and (e) in the SPP and PSS configurations, respectively. Each map is the average of 20 frames processed in Fiji and is displayed in camera counts using a common lookup table and intensity scale. No contrast adjustments were applied, and all maps share the 50-600-count intensity scale shown in panel (e). Scale bars in (b,d,e): 10 µm.

By changing the polarization configurations, we observe changes in the relative spatial contrast of the constituent regions across the same heterostructure (Figure 2d,e). Features prominent in one polarization projection are suppressed in the other, under identical acquisition settings. The cross-polarized measurements provide further support for sensitivity to out-of-plane components (Figure 2e). In Figures 1c,d and 3a, native hBN exhibits a stronger SSS than PPP response under identical acquisition conditions, providing a primarily in-plane reference. By contrast, the $V_{B^-}$ -rich region is most prominent in SPP, whereas the complete overlap remains intense in both SPP and PSS. In the high-numerical-aperture collection geometry used here, the cross-polarized SHG response is sensitive to out-of-plane nonlinear polarization alongside in-plane contributions, whereas SSS remains predominantly sensitive to the in-plane response. A substantial out-of-plane component has also been predicted for the $V_{B^-}$ optical transition [22, 23, 26, 27, 28]. Although the cross-analyzed field contains mixed components, its sensitivity to out-of-plane contributions provides the basis for mechanistic analysis developed below.

**Defects and interface coupling reshape the nonlinear anisotropy**

To determine the origin of spatial SHG contrast, we record angular polarization scans from four regions identified in Figure 2. Here, the incident linear polarization was rotated while the detection analyzer was fixed in either P or S polarization, and the resulting SHG intensities were plotted in polar coordinates (Figure 3). The angular response was described empirically by

$$I_{\mathrm{SHG}}(\theta) = c_0 + c_2\cos(2\theta) + s_2\sin(2\theta) + c_4\cos(4\theta) + s_4\sin(4\theta) \qquad \text{Eq.1}$$

where θ is the incident optical-polarization angle. The twofold and fourfold Fourier amplitudes were calculated as $A_2 = \sqrt{(c_2^2 + s_2^2)}$ and $A_4 = \sqrt{(c_4^2 + s_4^2)}$, respectively. This empirical decomposition quantifies the angular anisotropy of the measured intensity without assigning the Fourier components directly to individual elements of $\chi^{(2)}$ consistent with polarization-resolved analyses of layered nonlinear materials [32, 33, 42, 43].

The native-hBN region exhibits a weaker P-analyzed than S-analyzed response under identical acquisition conditions, consistent with the polarization contrast observed in Figure 1c,d (see Figure 3a). Both analyzer configurations produce predominantly four-lobed patterns, with their principal maxima separated by approximately 45° in incident-polarization angle. In a fixed-analyzer measurement, this four-lobed response represents a projection of the nonlinear polarization onto the laboratory detection axes and does not imply fourfold crystallographic symmetry. Such responses are compatible with a non-centrosymmetric hBN contribution with $D_{3h}$ symmetry under fixed-analyser detection [31, 32, 38] . This native-hBN response therefore provides an experimental reference against which irradiation- and overlap-induced changes can be evaluated. For irradiated-hBN region containing $V_{B^-}$ defects, we observe a pronounced imbalance between adjacent lobes (Figure 3b), with fitted ratios $A_2/A_4 = 0.98$ for P-analyzed detection and 1.15 for S-analyzed detection. The highest symmetry of an isolated $V_{B^-}$ centre is $D_{3h}$, but strain, local electric fields and Jahn-Teller distortions can lower its local symmetry and relax the associated optical selection rules [22, 23, 26]. Irradiation can additionally introduce spatially heterogeneous defect densities and strain. The enhanced twofold component is therefore consistent with irradiation-induced symmetry lowering arising from local strain, electric fields and spatially heterogeneous defect distributions [23, 26, 37].

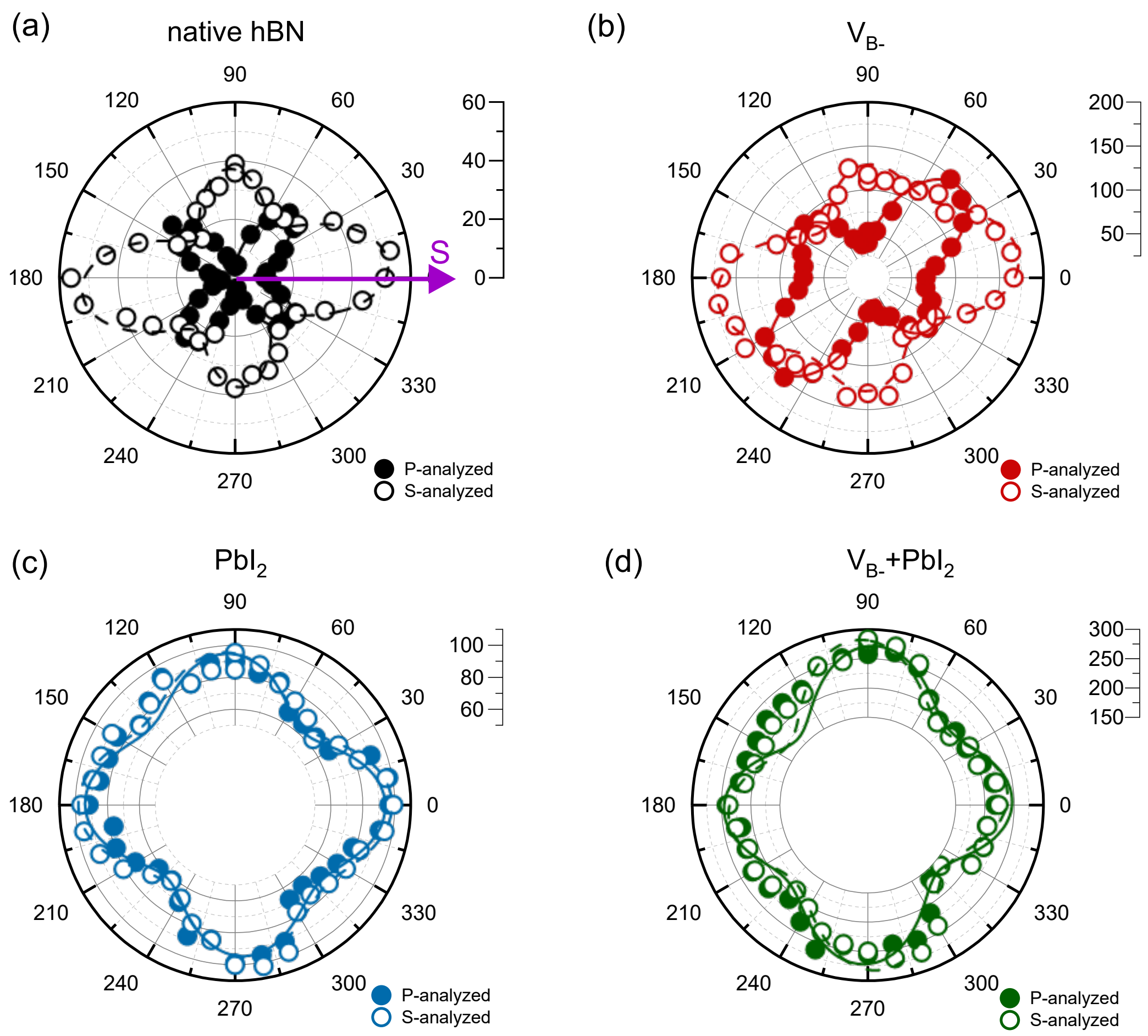

**Figure 3. Polarization-resolved SHG from four regions of the heterostructure.** SHG intensity as a function of incident linear-polarization angle, measured with the detection analyzer fixed in P or S polarization. Purple arrow indicates the S-polarization for incident beam. (a) Native hBN region, exhibiting a four-lobed angular pattern. (b) Neutron-irradiated hBN region containing negatively charged boron vacancies ($V_{B-}$), exhibiting a distorted angular pattern with a pronounced two-lobed character. (c) $PbI_2$ region, exhibiting a four-lobed pattern with approximately aligned angular maxima in the P- and S-analyzed channels. (d) Overlap region between neutron-irradiated hBN and $PbI_2$, exhibiting a balanced angular pattern distinct from the irradiated-hBN region. Filled and open circles denote P- and S-analyzed SHG intensities, respectively, with the same color used for both channels within each region. Symbols represent measured data and solid lines are fits to the empirical Fourier model defined in Eq. 1. Intensities are background-corrected mean camera counts extracted from equal-area ROIs. The incident-polarization angle is defined in the laboratory frame, with 0° and 180° corresponding to S-polarized incidence and 90° and 270° to P-polarized incidence. The auxiliary scales to the right indicate the mean camera-count scale used for each panel.

By contrast, the hBN/$PbI_2$ region without irradiated hBN (Figure 3c) retains a predominantly four-lobed modulation, with similar SHG intensity and nearly aligned principal maxima in the S- and P-analyzed channels (Figure 3c). Its mean SHG intensity is approximately 1.5 times that of native hBN under identical conditions. $PbI_2$ is a layered semiconductor with thickness- and environment-sensitive optical properties [44-47], and nonlinear microscopy has shown that its harmonic response varies with thickness, crystal orientation, substrate and excitonic resonance [48]. The modest increase can therefore reflect an additional $PbI_2$ or interface contribution, or coherent addition with the hBN-generated field [32, 33]. Interestingly, the complete hBN/$V_{B^-}$/$PbI_2$/hBN overlap, by contrast, shows a predominantly four-lobed response similar to $PbI_2$, with an approximately sixfold higher mean SHG intensity than the native-hBN region under the same conditions (Figure 3d). The fitted $A_2/A_4$ ratios are 0.19 for P-analyzed and 0.29 for S-analyzed detection. The overlap pattern is therefore not a uniformly amplified version of the irradiated-hBN response: both its angular envelope and analyzer dependence are changed. Coupling to $PbI_2$ therefore changes the relative nonlinear-dipole contributions and angular anisotropy of the defect-containing stack rather than merely amplifying the irradiated-hBN response. The origin of this enhanced overlap $I_{SHG}$ may include contributions from an altered second-order susceptibility, modified local fields and Fresnel factors, and coherent interference between SH fields generated within the different layers [32, 36, 38, 48]. Because the detected intensity is the squared magnitude of the coherently summed SH field, the Fourier amplitudes cannot be assigned directly to individual $\chi^{(2)}$ components without a multilayer model incorporating layer thicknesses, complex refractive indices, nonlinear tensors and relative phases [36, 38, 43], which is beyond the scope of present study.

**Interface-modified nonlinear polarization accompanies enhanced $V_{B^-}$ emission**

The native-hBN/$V_{B^-}$/$PbI_2$/native-hBN architecture (Figure 2a) places the optically active $PbI_2$ layer in direct contact with the defect-containing hBN, while preserving the encapsulated van der Waals interface (Figure 2a,b and Figure 4a,b). The room temperature photoluminescence, collected using excitation with a 488 nm laser and integrated around the $V_{B^-}$ emission band from 795 nm to 805 nm, is concentrated in the region where $PbI_2$ overlaps with irradiated hBN (Figure 4c). At room temperature, the peak $V_{B^-}$ emission from this region is enhanced by approximately 5-10-fold relative to irradiated hBN without $PbI_2$, consistent with the range of room-temperature photoluminescence enhancement reported for the same heterostructure architecture [28]. Mayner et al. attribute this behaviour to interlayer coupling between $PbI_2$ and $V_{B^-}$ via a photon ratcheting mechanism, with nonlinear photon-mediated excitation acting alongside other short-range energy-transfer pathways. In the present measurements, the corresponding four-layer combination produces the enhanced SHG intensity and pronounced change in nonlinear anisotropy identified in Figures 2 and 3. The $V_{B^-}$/$PbI_2$ interface architecture therefore supports both enhanced defect emission and a modified coherent second-order optical response.

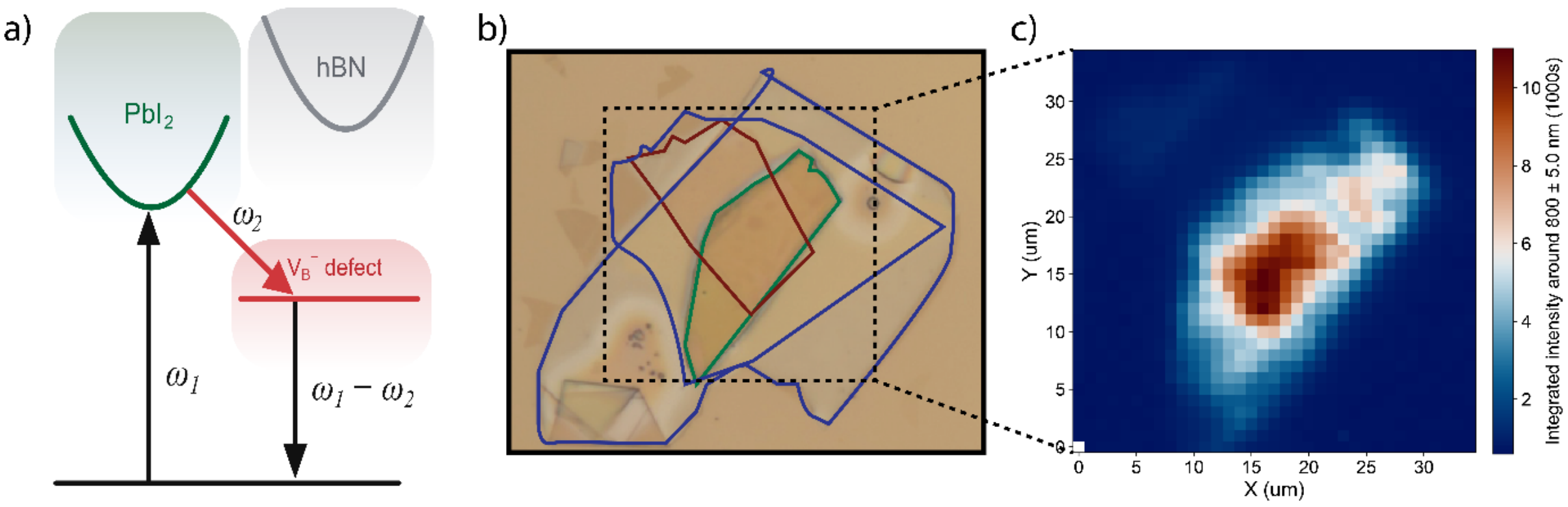

**Figure 4. Interfacial coupling links enhanced $V_{B-}$ emission to modified nonlinear polarization.** (a) Proposed photon-conversion mechanism, illustrating a pathway for nonlinear-optically mediated excitation of $V_{B-}$ defects involving $PbI_2$. Adapted from ref. [28]. (b) White-light image of an independently measured heterostructure with the same four-layer architecture; the dashed square indicates the mapped area. Enhanced PL intensity is observed in the region where $V_{B-}$ overlaps with $PbI_2$. The layers are outlined for clarity: The native hBN capping layers are in dark blue. $PbI_2$ is shown in green. Neutron irradiated hBN is shown in dark red. (c) Corresponding room temperature photoluminescence map integrated from 795 to 805 nm, showing enhanced emission in the irradiated-hBN/$PbI_2$ overlap. The excitation energy was 488 nm.

The photon-ratcheting mechanism proposed for the $V_{B-}$/$PbI_2$ interface by Mayner et al. [28] provides a microscopic framework for this behaviour (Figure 4a) [28]. Mayner et al. show that the optical response of the coupled system is described through the second-order susceptibility, with $PbI_2$-induced changes in emission traced to modifications of the underlying electronic structure and optical matrix elements. In their model, the mixed-frequency photon-ratcheting pathway probes $\chi^{(2)}(-\Omega;\omega_1,-\omega_2)$, where $\Omega=\omega_1-\omega_2$ corresponds to the defect excitation energy. The present study probes the degenerate second-harmonic response $\chi^{(2)}(-2\omega;\omega,\omega)$. The same formalism used for the photon-ratcheting calculation explicitly reduces to the standard SHG response at $2\omega$. The two measurements therefore access distinct frequency channels of the quadratic susceptibility of the same coupled interface.

In the proposed photon-ratchet model, the resonant perturbation chain contains mixed optical matrix elements connecting a defect-localized state in hBN with $PbI_2$ electronic states, so that the enhancement originates at the interface rather than from a remotely propagating nonlinear field. The corresponding nonlinear channel does not require the formation of a real exciton population in $PbI_2$; instead, its source involves $PbI_2$ states within the first one or two layers adjacent to the interface. At the 950-nm fundamental used here, the generated second harmonic at 2.61 eV lies within the $PbI_2$ electronic-transition regime [44, 48]. The pronounced enhancement together with the altered polarization symmetry is therefore interpreted as an interface-induced modification of the optical matrix elements contributing to $\chi^{(2)}$, rather than as simple population transfer or uniform local-field amplification. Such a modification does not require strong ground-state hybridization. The calculations show that the electronic and spin structure of $V_{B^-}$ remains largely preserved upon coupling to $PbI_2$, while the nonlinear optical response can nevertheless change strongly. This is possible because $\chi^{(2)}$ depends on products of optical transition matrix elements and resonant energy denominators, which allows weak interlayer coupling to strongly modify the nonlinear response without any substantial ground-state hybridization. The large SHG enhancement and redistribution of SPP and PSS contributions are therefore compatible with weakly hybridized defect states whose transition amplitudes are nevertheless strongly modified by the interfacial environment.

The occurrence of enhanced $V_{B^-}$ photoluminescence and the reconstructed SHG response in the same layer architecture links two complementary manifestations of the coupled interface. Where the photon-ratcheting measurements establish enhanced defect excitation and emission, polarization-resolved SHG independently reveals modification of the underlying quadratic optical response. Together with the microscopic $\chi^{(2)}$ framework developed for the same heterostructure [28], these observations support an interfacial picture in which $PbI_2$-$V_{B^-}$ coupling modifies resonant optical matrix elements without requiring substantial ground-state hybridization. Negatively charged boron-vacancy defects therefore act not only as acceptors in interlayer energy-transfer processes but also as an optically active component of the effective second-order susceptibility of the van der Waals heterostructure.

## Conclusions

Polarization-resolved SHG imaging reveals that heterostructure formation modifies the nonlinear optical response of hBN in distinct ways. The quadratic pulse-energy dependence is consistent with a dominant second-order response, while analyzer-resolved angular scans show that $V_{B^-}$-rich hBN develops a strong twofold distortion relative to native hBN. In the complete $V_{B^-}/PbI_2$ overlap, the response returns towards a predominantly fourfold pattern while the SHG intensity increases by approximately sixfold relative to the native-hBN region. The occurrence of this reconstructed coherent response in the same layer architecture previously shown to enhance $V_{B^-}$ photoluminescence identifies the interface architecture as the common origin of both effects. Comparison with the independently developed $\chi^{(2)}$ framework for the same heterostructure supports an interface-modified quadratic susceptibility without requiring strong ground-state hybridization. These results establish polarization-resolved SHG as a complementary probe of defect-mediated interlayer optical coupling in van der Waals heterostructures.

## Acknowledgments

L.D. and P.S. would like to acknowledge Prof. Sylvie Roke (LBP, EPFL) and Prof. Aleksandra Radenovic (LBEN, EPFL) for the discussions. E.S.M. would like to thank CMI EPFL for the technical support and facilities. L.D. and P.S. thank the Julia Jacobi Foundation, the Swiss National Science Foundation (Grant 200021-182606-1), and the European Research Council (R2-tension 951324) for financial support.

## Supporting Information

## S1. Methods and materials

### S1.1 Heterostructure Fabrication

Commercially available hBN and $PbI_2$ bulk crystals were acquired from HQ Graphene. The hBN bulk crystal was then neutron irradiated at facilities at OSU to generate $V_{B^-}$ defects. A dose of $2.6 \times 10^{16}$ n cm$^{-2}$ was used as previously described in Ref. [49]. The heterostructure was fabricated by a dry-transfer technique using polycarbonate (PC) membranes. Individual flakes were mechanically exfoliated onto $SiO_2$/Si substrate with oxide thickness of 270 nm using the cleavage technique method pioneered for the production of graphene [50]. The same mechanical exfoliation technique was used for $PbI_2$ and hBN that had not been neutron irradiated. The heterostructure was made by picking up the flakes successively using the PC stamp and then releasing them onto ODMR antennas by increasing the temperature above 200 °C, causing the PC to pass through its glass-transition temperature and remain on the target substrate [51]. The residual PC was subsequently removed using chloroform, followed by isopropanol. Gel-Pak and $SiO_2$/Si substrates containing exfoliated flakes were stored in an argon glovebox between fabrication steps. The thickness of the individual layers was not controlled during exfoliation. The measured thickness of $V_{B^-}$ layer is 4 nm and $PbI_2$ is 180 nm.

### S1.2 Second Harmonic microscopy

The microscope excitation source consisted of a femtosecond laser system comprising a Carbide laser (80 W, 200 μJ, 400 kHz; Light Conversion) pumping an optical parametric amplifier (ORPHEUS-NEO; Light Conversion). The system delivered pulses with durations of <35 fs at a repetition rate of 400 kHz and is continuously tunable over the wavelength range of 640-1300 nm, providing an average output power of >4 W at 950 nm. The output beam was subsequently passed through a pulse compressor (Light Conversion Ltd.).

For second harmonic-generation (SHG) imaging, the excitation wavelength was set to 950 nm. The pulses were negatively chirped by approximately −2000 fs² to compensate for dispersion in the optical system and obtain a pulse duration of approximately 35 fs at the sample plane. A lens with a focal length of 20 cm (Thorlabs) focused the excitation beam onto the back focal plane of a 60× water-immersion objective (LUMPFLN 60XW, NA 1.1; Olympus). This configuration produced wide-field illumination with a diameter of approximately 90 μm at the sample plane, with the excitation beam incident nominally normal to the sample surface. For the measurements shown in Figures 1-3, the average excitation power at the sample plane was set to 40 mW. At a repetition rate of 400 kHz, this corresponds to a pulse energy of 0.1μJ. For a circular illumination area with a diameter of 90 μm, the corresponding fluence was approximately 1.57 mJ $cm^{-2}$. For a pulse duration of 35 fs, the estimated peak intensity was approximately 44.9 GW $cm^{-2}$.

The generated SHG signal was collected in reflection geometry through the same 60× water-immersion objective used for excitation (epi-detection configuration). The collected SHG signal was imaged by a single-photon avalanche diode (SPAD) camera (PI Imaging, SPAD 512) using an 18 cm tube lens. A 470 nm bandpass filter (FBH470-10; Thorlabs) was placed in the detection path to suppress residual fundamental light at 950 nm. The camera acquisition was synchronized with the laser repetition rate. SHG images were acquired using an integration time of 1s per image, with 20 frames accumulated or averaged for each measurement. The SPAD camera was operated in gated mode.

Bright-field imaging was performed using an LED source (M660L4; Thorlabs). The bright-field signal was spectrally separated from the SHG detection path using a 650 nm short-pass dichroic mirror (DMSP650; Thorlabs) and detected using a separate CMOS camera (CS165MU1/M; Thorlabs). This configuration enabled simultaneous acquisition of SHG and bright-field images.

The microscope was constructed on a Rapid Automated Modular Microscope platform (RAMM; Applied Scientific Instrumentation, ASI). Sample positioning was controlled using an XYZ translation stage (PZ-2000FT; ASI). The X and Y axes were driven by motors equipped with linear encoders, whereas axial positioning was controlled by a piezoelectric actuator with a travel range of 300 μm.

### S1.3 Polarization control and acquisition sequence

A Glan-Taylor polarizer and polarizing beam splitter defined the input polarization, and an achromatic half-wave plate rotated the incident polarization. The half-wave plate was advanced in 5° mechanical steps, corresponding to 10° steps in optical polarization. The laboratory angular offset was removed before plotting, such that 0° and 180° denote S-polarized incidence and 90° and 270° denote P-polarized incidence.

The detection analyzer was held fixed in either P or S while the incident polarization was rotated. For P-analyzed detection, 90° samples the PPP projection and 0° samples PSS. For S-analyzed detection, 0° samples SSS and 90° samples SPP. The first letter identifies the detected SH polarization, and the final two letters identify the incident fundamental polarization. Because the beam is delivered and collected by a high-numerical-aperture objective, P and S denote the laboratory basis of the optical train rather than a single plane-wave Fresnel basis for every ray.

Each SHG image was acquired with a 1 s integration time. Twenty consecutive frames were averaged for visualization and quantitative analysis.

**S1.4 Image processing and analysis**

SHG images were processed and analysed using Fiji/ImageJ (ImageJ version 1.54t, National Institutes of Health, Bethesda, MD, USA). Unless stated otherwise, identical processing parameters were applied to all images included in a given comparison. For each measurement, 20 consecutive frames were acquired and subsequently averaged in Fiji/ImageJ to produce the final SHG image used for visualization and quantitative analysis. Regions of interest were defined using segmentation and the mean SHG intensity within each region was calculated after subtraction of the background signal. Any adjustments to brightness, contrast, or lookup tables were applied only for visualization and did not affect the quantitative analysis.

**S2. Polarization convention and angular analysis**

**S2.1 Effective second-order response**

Within the electric-dipole approximation, the coherent nonlinear polarization generated at twice the incident frequency is

$$P_i(2\omega) = \varepsilon_0 \sum_{j,k} \chi^{(2)}_{ijk}(-2\omega;\omega,\omega)E_j(\omega)E_k(\omega) \qquad \text{Eq.2}$$

For a fixed analyzer channel q, the detected intensity can be expressed in an effective form as

$$I_{2\omega,q}(\theta) = K_q \left|\chi^{(2)}_{\mathrm{eff},q}(\theta)\right|^2 I_\omega^2, \qquad \text{Eq.3}$$

Here $K_q$ contains illumination, reflection, collection, transmission and detector factors that are not absorbed into the effective susceptibility. The quantity $\chi^{(2)}_{\mathrm{eff},q}$ is a polarization projection of the full response of the multilayer stack. Quantitative extraction of individual tensor elements requires calibrated local fields, Fresnel factors, crystal orientations and a vectorial high-numerical-aperture collection model.

For a fixed-analyzer angular response, the incident field in the laboratory polarization basis is

$$E(\theta) = E_p\cos(\theta) + E_s\sin(\theta) \qquad \text{Eq.4}$$

Because the nonlinear polarization is quadratic in the incident field, the second-harmonic field projected onto a fixed analyzer can be expressed as

$$E_{2\omega,q}(\theta) = a_q + b_q\cos(2\theta) + d_q\sin(2\theta) \quad \text{Eq.5}$$

where the generally complex coefficients include the effective nonlinear response and the optical transfer function. Because the nonlinear polarization is quadratic in the incident field, its projected field contains constant, cos 2θ and sin 2θ terms. Squaring the projected field generates intensity components at 0, 2θ and 4θ. A predominantly four-lobed fixed-analyzer pattern is therefore consistent with a three-fold hBN nonlinear tensor and does not imply fourfold crystallographic symmetry. By contrast, co-rotating parallel or crossed polarization measurements of an ideal non-centrosymmetric $D_{3h}$ layer produce the familiar six-lobed SHG intensity pattern. Linear-polarization orientation is unchanged by a 180° rotation. The measured intensity must therefore satisfy I(θ + 180°) = I(θ).

### S2.2 Parametric fitting

To compare the polarization-dependent SHG response across the sample, we selected a region of interest (ROI) within each distinct sample region and extracted its mean SHG intensity at each recorded excitation polarization angle for both fixed analyzer orientations, P and S. The resulting angular dependences were fitted using a truncated Fourier expansion containing second- and fourth-order angular terms given in Eq. 1. Here, θ denotes the incident linear-polarization angle relative to the experimental reference axis, expressed in radians. The coefficient ($c_0$) represents the angular mean of the fitted intensity, while the coefficient pairs ($c_2$, $s_2$) and ($c_4$, $s_4$) describe the twofold and fourfold angular components, respectively. Including both sine and cosine terms allows the orientation of each component to vary relative to the reference axis. Their amplitudes are defined as

$$A_2 = \sqrt{c_2^2 + s_2^2}, \qquad A_4 = \sqrt{c_4^2 + s_4^2}. \qquad \text{Eq.6}$$

A larger ratio ($A_2/A_4$) indicates a greater relative contribution of the twofold component to the measured angular pattern. This phenomenological analysis quantifies the angular anisotropy without directly assigning a crystallographic point-group symmetry.

**Table S1. Fit coefficients for each of the four designated regions**

| | **P-analysed** | | | **S-analysed** | | |
|---|---|---|---|---|---|---|
| **ROI** | **$A_2$** | **$A_4$** | **$A_2/A_4$** | **$A_2$** | **$A_4$** | **$A_2/A_4$** |
| **native hBN** | 3.27 | 105.78 | 0.03 | 60.48 | 138.68 | 0.43 |
| **$V_{B-}$** | 1015.86 | 1033.99 | 0.98 | 832.33 | 721.11 | 1.15 |
| **$PbI_2$** | 20.89 | 183.02 | 0.11 | 28.61 | 168.15 | 0.17 |
| **Overlap** | 310.82 | 1636 | 0.19 | 505.15 | 1730.61 | 0.29 |

**Author contributions**

L.D. and P.S. performed experiments. L.D. and P.S. developed the experimental and analytical workflow. P.S. and L.D. interpreted the experimental data. P.S. and L.D. wrote the manuscript with input from all the authors. E. S. M. prepared the hBN flakes and vdW heterostructures used in this study. P.S. conceived and supervised the work.

**Competing interests**

The authors declare no competing interests.